\documentstyle[aps,prl,epsfig]{revtex}
\begin{document}
\draft

\title
{Two component interacting bosons: 1-d exact results}

\author{You-Quan Li$^{1,2}$, Shi-Jian Gu$^1$, 
Zu-Jian Ying$^1$, and Ulrich Eckern$^2$}

\address{
$^1$Zhejiang Institute of Modern Physics, Zhejiang University, 
    Hangzhou 310027, P.R. China\\
$^2$Institute for Physics, Augsburg University, D-86135 Augsburg, Germany
}

\date{Received: June 29, 2000}

\maketitle

\begin{abstract}
Motivated by the experiment of two-component Bose-Einstein condensates
produced in magnetically trapped $^{87}Rb$, 
we study one dimensional Boson systems with repulsive $\delta$-function 
interaction in the presence of SU(2) intrinsic degree of freedom
by means of coordinate Bethe ansatz. 
The ground state and low-lying excitations are solved by
both numerical and analytical methods. It is shown that the ground
state is an isospin-ferromagnetic state and the excitations are composed
of three elementary particles: holons, antiholons and isospinons.
The isospinon is a triplet coupled to the ``ferromagnetic'' background
antiparallelly.
\end{abstract}

\nopagebreak
\twocolumn

Exactly savable models 
\cite{WiegmannAP83,AndreiEMP83,Haldane83,deVegaNP85,%
KorepinBook,LiebW,LiebL,Gaudin67,Yang67,Sutherland68,%
ChoyHaldane,Schlottman,Gaudin71,Schulz,Woynarovich85,LiG98}
are known to play an important
role in the investigation of one-dimensional interacting many-particle 
systems. These approaches have served as a great source of inspiration
for the new understandings of non-perturbative phenomena of correlated 
electronic systems, for example, the spinon was explicitly demonstrated
on the basis of one-dimensional exact solution of Hubbard model \cite{LiebW}.
Among those models, an earlier typical one is the one-dimensional bosons
with repulsive $\delta$-function interaction which was solved \cite{LiebL}
by means of Bethe-ansatz method. The method was developed to solve
the problem of spin-1/2 fermions \cite{Gaudin67,Yang67} with 
$\delta$-function interaction. Since then had been there various extensions,
such as the consideration of electrons on crystalline lattice \cite{LiebW},
the generalization to the systems with higher symmetries
\cite{Sutherland68,ChoyHaldane,Schlottman} and applications to different
boundary conditions \cite{Gaudin71,Schulz,Woynarovich85,LiG98}.
Nevertheless, the features of SU(2) bosons with $\delta$-function interaction,
as far as we are aware, has not been studied hitherto. 
Because particles with ``spin-1/2'' are regarded as fermions,
the two-component model with anti-symmetric permutation was
studied thoroughly but the symmetric case did not absorb much attention. 
Recently, however, a two component Bose gas was produced in magnetically
trapped $^{87}$Rb by rotating the two hyperfine states
into each other with the help of slightly detuned Robi oscillation field
\cite{WilliamsTH,WilliamsEXP}. 
It was noticed \cite{Janson} that the ground state for a bosonic system can
be surprisingly different from the traditional scalar Bose system once 
acquiring intrinsic degree of freedom where variational method was used.
It is therefore very important to study the features of the ground state
and low-lying excitations of the multi-component interacting Bose systems.

In order to attain non-perturbative insight on the features of such system,
in present letter we study the model of one-dimensional
SU(2) bosons with repulsive $\delta$-function interaction which is an
integrable model.
Showing the connection with the coupled Gross-Pitaevski
equation \cite{Gross,Pitaevski} related to Boson Einstein 
condensation \cite{WilliamsTH}, we first solve the Bethe-ansatz
equation for bosons with SU(2) intrinsic degree of freedom. 
The distribution feature of the quasi-momentum determined by the 
Bethe-ansatz equation is investigated in week coupling limit. 
The ground state property and low-lying excitations are studied by 
both numerical calculation and thermodynamic limit approach.  
We fascinatedly found that the ground state is not an ``anti-ferromagnetic'' 
but a `ferromagnetic' state, the charge-isospin phase separation exists 
and the isospinon is a triplet in stead of doublet which is well known
for spinons in spin-1/2 Fermi system.

The two-component Bose gas produced in a recent experiment is known 
to satisfy the coupled  Gross-Pitaevski equations:
\[
i\hbar\frac{\partial}{\partial t}
 \pmatrix{\psi_1 \cr \psi_2}=
  \pmatrix{\hat{G} & \hat{P}^* \cr \hat{P} & \hat{G}}
   \pmatrix{\psi_1 \cr \psi_2 }.
\]
where 
$\hat{G}=-\frac{\hbar^2}{2m}\nabla^2 + V({\bf r}) 
 +c\sum_a |\psi_a|^2$, $\hat{P}=\hbar\Omega/2$. 
$\Omega$ is Rabi oscillation field and $V({\bf r})$
the trapping potential. 
 
Considering the system trapped in a one dimensional ring of length $L$
and introducing 
$\phi_1=(\psi_1+\psi_2)/\sqrt{2}$, $\phi_2=(\psi_1-\psi_2)/\sqrt{2}$,
we obtain from the above Gross-Pitaevski equation for real $\Omega$ 
the following Hamiltonian: 
\[
H=\int dx \Bigl[\partial_x\phi_a^*\cdot\partial_x\phi_a
    +c\phi^*_a\phi^*_b\phi_b\phi_a 
     -(-1)^a \Omega\phi^*_a\phi_a\Bigr] 
\]
where $a,b=1,2$, the repeated indices denote summation and 
natural unit are adopted for simplicity.
The fields obey bosonic commutation relations
$
[\,\phi^*_a(x),\,\phi_b(y)\,]
=\sum_{n\in{Z\!\!\!Z}}\delta_{a b}\delta(x-y-nL).
$
The Rabi oscillation field contributes a Zeemann-type term.
To avoid unnecessary confusions away from the conventional spins
which distinguish between fermions and bosons in
the known spin-statistical connection theorem, 
we denote the generators of isospin SU(2) by ${\cal I}$, correspondingly  
$
[{\cal I}^+, {\cal I}^-]=2{\cal I}^z.
$
The intrinsic degree of freedom 
can be specified by the eigenvalues of ${\cal I}^z$
or labeled by isospin up and down.

The Bethe-ansatz equations for two-component bosons
are  obtained as follows.
\begin{eqnarray}
e^{ik_jL}=-\prod^{N}_{l=1}\Xi_1(k_j-k_l)
   \prod_{\nu=1}^{M}\Xi_{-1/2}(k_j-\lambda_\nu)
    \nonumber\\
1=-\prod^{N}_{l=1}
    \Xi_{-1/2}(\lambda_\gamma-k_l)
     \prod^{M}_{\nu=1}
      \Xi_1(\lambda_\gamma-\lambda_\nu)
\label{eq:BAE}
\end{eqnarray}
where $\Xi_\beta(x)=(x+i\beta c)/(x-i\beta c)$.
Eq.(\ref{eq:BAE}) determines the values of the 
quasi-momenta $\{k_j\}$ and the
isospin rapidities $\{\lambda_\nu\}$ for a $N-2M+1$ fold multiplet
characterized by the total isospin ${\cal I}_{tot}=(N-2M)/2$.
Eq. (\ref{eq:BAE}) is mimic to the Bethe ansatz equation of BBB case
of Ref. \cite{Sutherland75}
except a variant in the exponential.

Equation (\ref{eq:BAE}) is obtained by the following procedure.
Applying the Hamiltonian 
to the Hilbert space of $N$ particles and considering its first quantized
version on the domain ${\sf l\!\!R}\setminus\{{\sf l\!\!P}_{ij}\}$ where
${\sf l\!\!P}_{ij}:=\{x|x_i-x_j=0\}$ is the hyperplane defined by
the $\delta$-function singularity,
we get that only the $N$-dimensional Laplace operator remains in the
Schr\"{o}dinger operator.
Thus $N$-dimensional plane waves solve the Schr\"{o}dinger equation.
We sum up all the plane waves of which the wave vectors are just
permutations of a definite  ${\bf k}=(k_1, k_2,\ldots,k_N)$
according to the Bethe ansatz strategy.
Integrating the Schr\"{o}dinger equation across the hyperplanes, we obtain 
$S(k_i-k_j)=[k_i-k_j-ic{\cal P}]/[k_i-k_j+ic]$, which connects the 
Bethe-ansatz wave functions between the regions separated by the 
hyperplanes and $\check{S}:={\cal P}S$ (${\cal P}$ stands for the spinor
representation of the permutation group $S_N$) relates the coefficients of 
different plane waves in the same region. 
It is worth to point out that the bosonic permutation symmetry (instead of 
the antisymmetry) was imposed when solving the $S$-matrix.
Analogous to the case of spin-1/2 fermions \cite{Yang67},
the periodic boundary condition leads to an eigen-equation 
for the product of the $S$-matrix. 
As the obtained $S$-matrix satisfies the Yang-Baxter
equation, the quantum inverse scattering method \cite{Faddeev} is
applicable for diagonalization. 
After writing out the fundamental commutation relations and evaluating
the eigenvalues of the reference state 
$|\omega\rangle=|\uparrow\uparrow\ldots\uparrow\rangle$,
one may immediately realize the difference from that 
of spin-1/2 fermions. For example, 
$A(\xi)|\omega\rangle=\prod_l (\xi-\xi_l-ic)/(\xi-\xi_l+ic)|\omega\rangle$,
$D(\xi)|\omega\rangle=\prod_l (\xi-\xi_l)/(\xi-\xi_l+ic)|\omega\rangle$
in the notion of ref.\cite{Faddeev,Kulish}.
Consequently, we got the result of eq. (\ref{eq:BAE}).
It is an facilitating picture that the Bethe ansatz strategy implies 
the existence of infinitely many constants of motion,
$\sum_j k^n_j=constant$, in addition to the usual energy
$E=\sum_{l=1}^N k_l^2+\Omega(N-2M)$ and momentum $P=\sum_l^N k_l$.

The logarithm of eq. (\ref{eq:BAE}) gives rise to
\begin{eqnarray}
k_j=\frac{2\pi}{L}I_j+\frac{1}{L}\sum^{N}_{l=1}\Theta_1(k_j-k_l)
     &+&\frac{1}{L}\sum_{\nu=1}^{M}\Theta_{-1/2}(k_j-\lambda_\nu),
     \nonumber\\
2\pi J_\gamma=\sum^{N}_{l=1}\Theta_{-1/2}(\lambda_\gamma-k_l)
  &+&\sum^{M}_{\nu=1}\Theta_1(\lambda_\gamma-\lambda_\nu),
\label{eq:logBAE}
\end{eqnarray}
where $\Theta_\beta(x):=-2\tan^{-1}(x/\beta c)$, 
and both the quantum numbers $I_j$ and $J_\gamma$ take $integer$ or 
$half$ $integer$ values depending on whether $N-M$ is $odd$ or $even$.
As a comparison, the Bethe-ansatz equation for 
spin-1/2 fermions is not only lack of the first summation 
but also has an opposite sign in the second summation 
in the first line of eq. (\ref{eq:logBAE}). 
The momentum is easily obtained from eq. (\ref{eq:logBAE}),
$P=\sum_l k_l=(\sum_l I_l-\sum_\nu J_\nu)2\pi/L$. 

It is inspirable to analysis eq. (\ref{eq:logBAE}) 
in the strong and weak coupling regimes.
For strong interaction $c\rightarrow\infty$ the wave function
vanishes for any $x_i=x_j$ and hence the bosons avoid each other
like fermions which is in agreement with the discussion of 
quantum degeneracy in trapped 1D gases \cite{Petrov}.

Due to $\Theta_1(x)\rightarrow-\pi{ \rm sgn}(x)$
and $\Theta_{-1/2}(x)\rightarrow\pi{ \rm sgn}(x)$ for $x\gg 1$,  
eq. (\ref{eq:logBAE}) for weak coupling limit $c\rightarrow 0$ becomes,
\begin{eqnarray}
k_j+\frac{\pi}{L}\sum_{\l=1}^N{\rm sgn }(k_j-k_l)
   -\frac{\pi}{L}\sum_{\nu=1}^M{\rm sgn }(k_j-\lambda_\nu)
     =\frac{2\pi}{L}I_j,
       \nonumber\\
\sum_{l=1}^N {\rm sgn }(\lambda_\gamma-k_l)
  -\sum_{\nu=1}^M{\rm sgn }(\lambda_\gamma-\lambda_\nu)
    =2J_\gamma.\;\;
\label{eq:weakBAE}
\end{eqnarray}
The subscript of the isospin rapidity $\lambda_\gamma$ can be chosen
in such a way that $J_\gamma$ is ranged in an increasing order, then
the second equation of eq. (\ref{eq:weakBAE}) turns to
\begin{equation}
\sum_{l=1}^N{\rm sgn }(\lambda_\gamma -k_l)
=2J_\gamma+2\gamma-M-1.
\label{eq:2weakBAE}
\end{equation}
Because $|J_\gamma|<(N-M+1)/2$ for a given $M$
and $M\leq N/2$ due to the restriction by Young tableau, the minimum
value of the right hand side of eq. (\ref{eq:2weakBAE}) is $-N+2$.
This requires that the smallest $k_l$ must be smaller than the smallest
$\lambda_\nu$, otherwise the left hand side would be $-N$.
Eq. (\ref{eq:2weakBAE}) also implies  
\begin{eqnarray}
\sum_{l=1}^N[{\rm sgn}(\lambda_{\gamma+1}-k_l)
  -{\rm sgn}(\lambda_\gamma-k_l)] \nonumber\\
=2(J_{\gamma+1}-J_\gamma+1).\;\;\;
\end{eqnarray}
Thus, for $J_{\gamma+1}-J_\gamma=m$, there must exist exactly $m+1$
solutions of $k_l$ satisfying $\lambda_\gamma<k_l<\lambda_{\gamma+1}$.
Furthermore, from the first equation of eq. (\ref{eq:weakBAE}) we obtain
\begin{eqnarray}
k_{j+1}&-&k_j 
  -\frac{\pi}{L}\sum_{\nu=1}^{M}
    [{\rm sgn}(k_{j+1}-\lambda_\nu)
     -{\rm sgn}(k_j-\lambda_\nu)]\nonumber\\
\,&=&\frac{2\pi}{L}(I_{j+1}-I_j -1).
\label{eq:1weakBAE}
\end{eqnarray}
Obviously, for $I_{j+1}-I_j=n$, there will be 
$k_{j+1}-k_j=2n\pi/L$ if there is a $\lambda_\gamma$
such that $k_j<\lambda_\gamma<k_{j+1}$, otherwise 
$k_{j+1}-k_j=(n-1)2\pi/L$. So an isospin rapidity of value
$\lambda_\mu$ always 
repels the quasi-momenta away from that value.
As a result, an existing $\lambda_\mu$ will suppresses the density of 
state in $k$-space at the point $k=\lambda_\mu$.
The more isospin rapidities there are, the higher the energy 
will be.
Thus the ground state of SU(2) interacting bosons is no more a 
SU(2) singlet but an isospin ``ferromagnetic'' state
which differs from Fermi case greatly. 

\begin{figure}
\epsfclipoff
\fboxsep=0pt
\setlength{\unitlength}{0.8mm}
\begin{picture}(76,42)(0,0)
\linethickness{1pt}
\epsfysize=36mm
\put(-1,1){{\epsffile{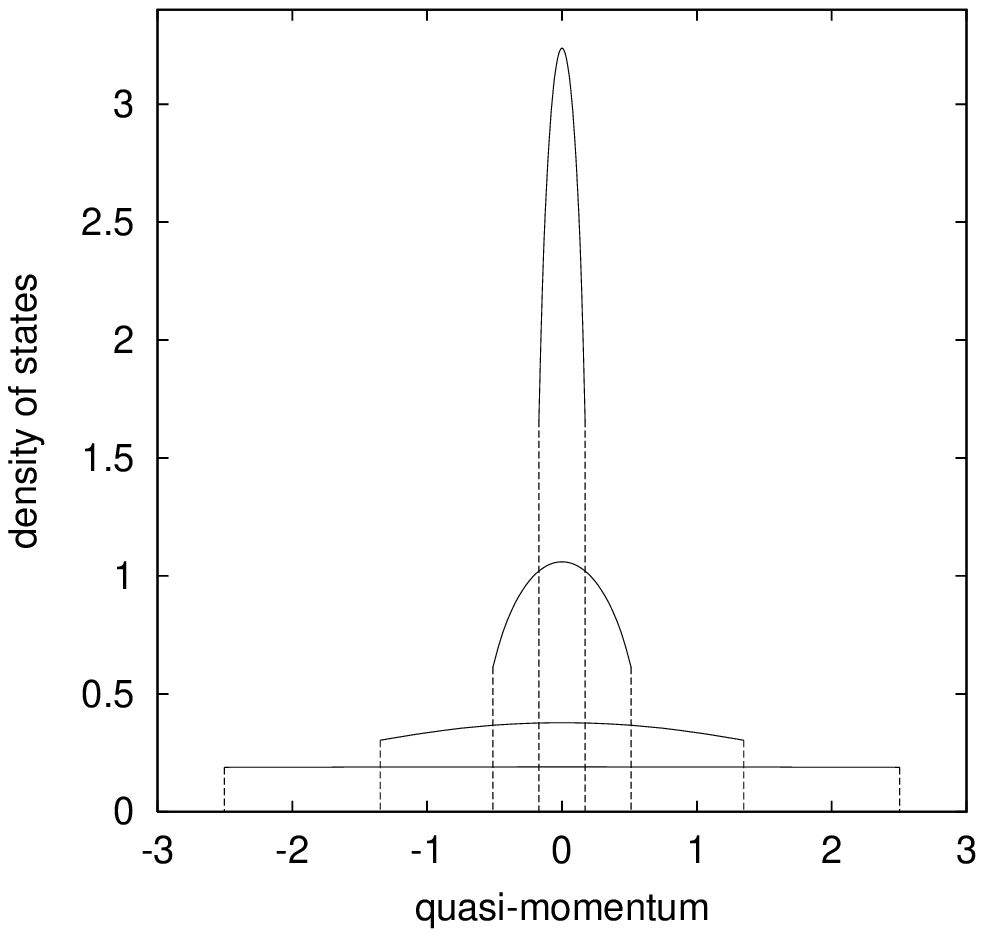}}}
\epsfysize=36mm
\put(49,1){{\epsffile{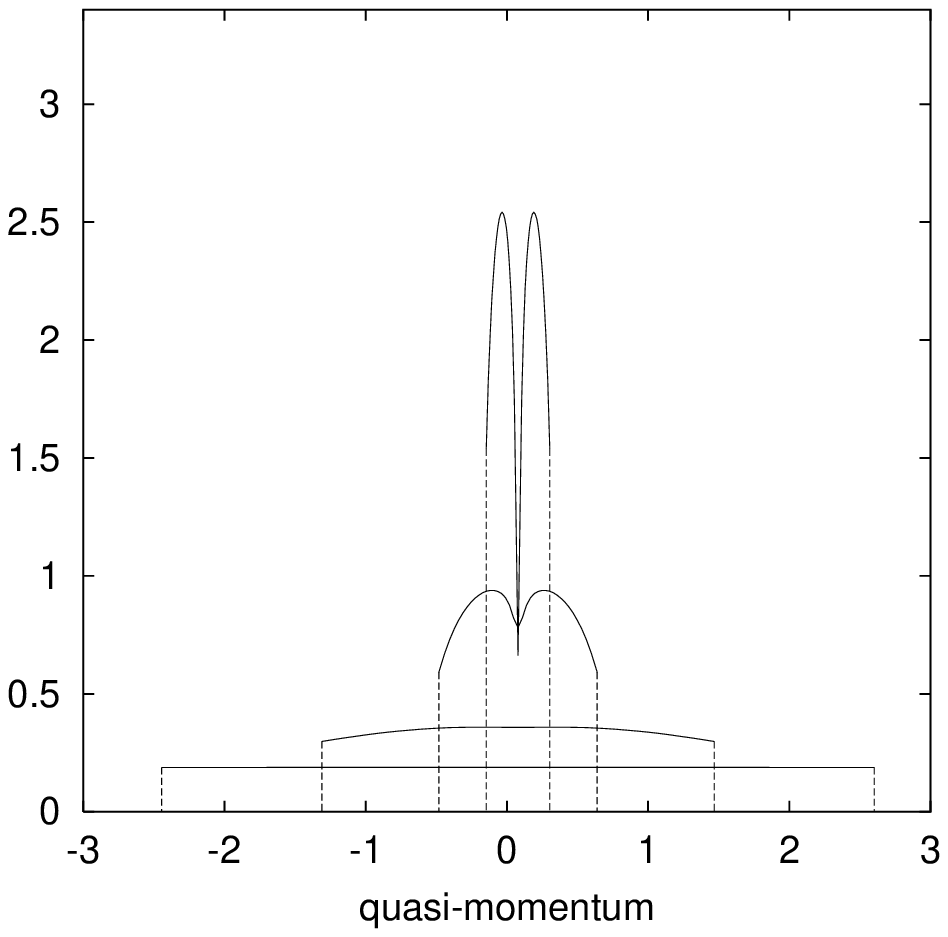}}}
\end{picture}
\vspace{0mm}
\caption{
The density of state per length
in $k$-space for the ground state (left)
and for the state in the presence of one isospin rapidity $\lambda$
by choosing $J_1=0$ (right).
The distribution changes from a histogram to a narrow peak gradually for the
coupling from strong to week $c=10, 1, 0.1, 0.01$.
The calculation is made for $L=40$, $N=40$.
The left panel is mimic to the fig.2 of ref [7].}
\label{fig:density}
\end{figure}

\begin{figure}
\epsfclipoff
\fboxsep=0pt
\setlength{\unitlength}{0.8mm}
\begin{picture}(76,42)(0,0)
\linethickness{1pt}
\epsfysize=36mm
\put(-1,1){{\epsffile{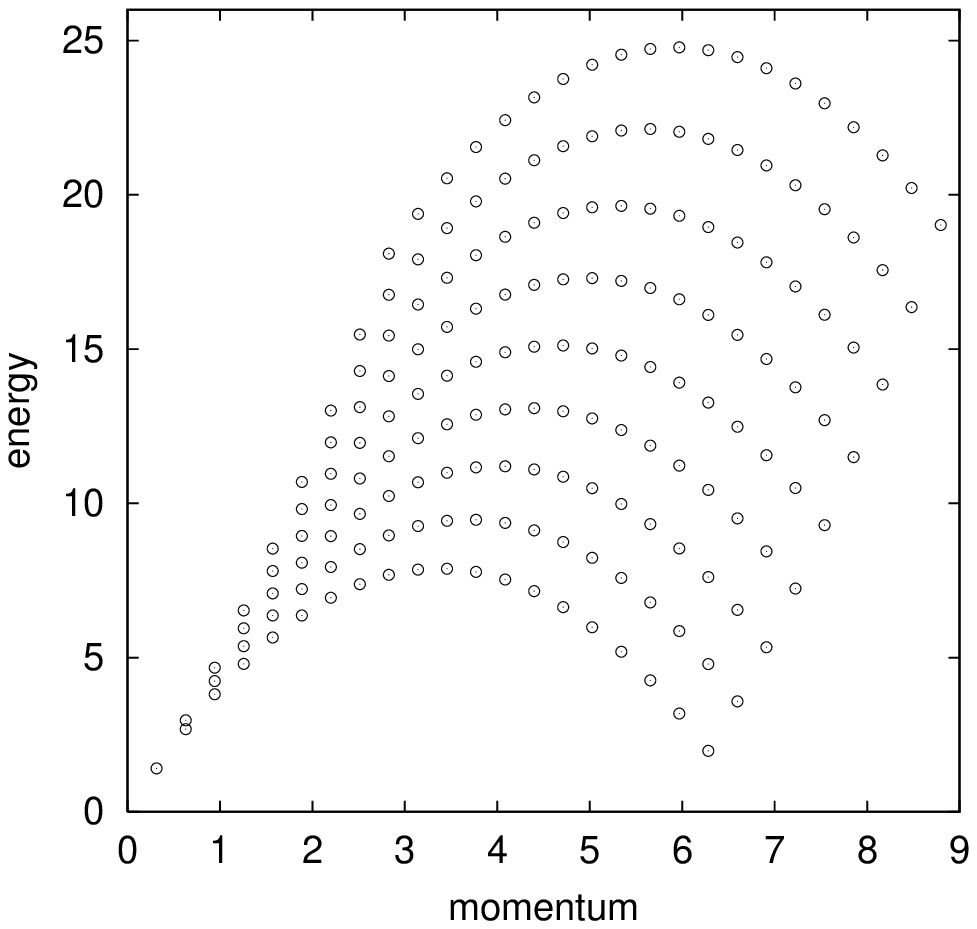}}}
\epsfysize=36mm
\put(50,1){{\epsffile{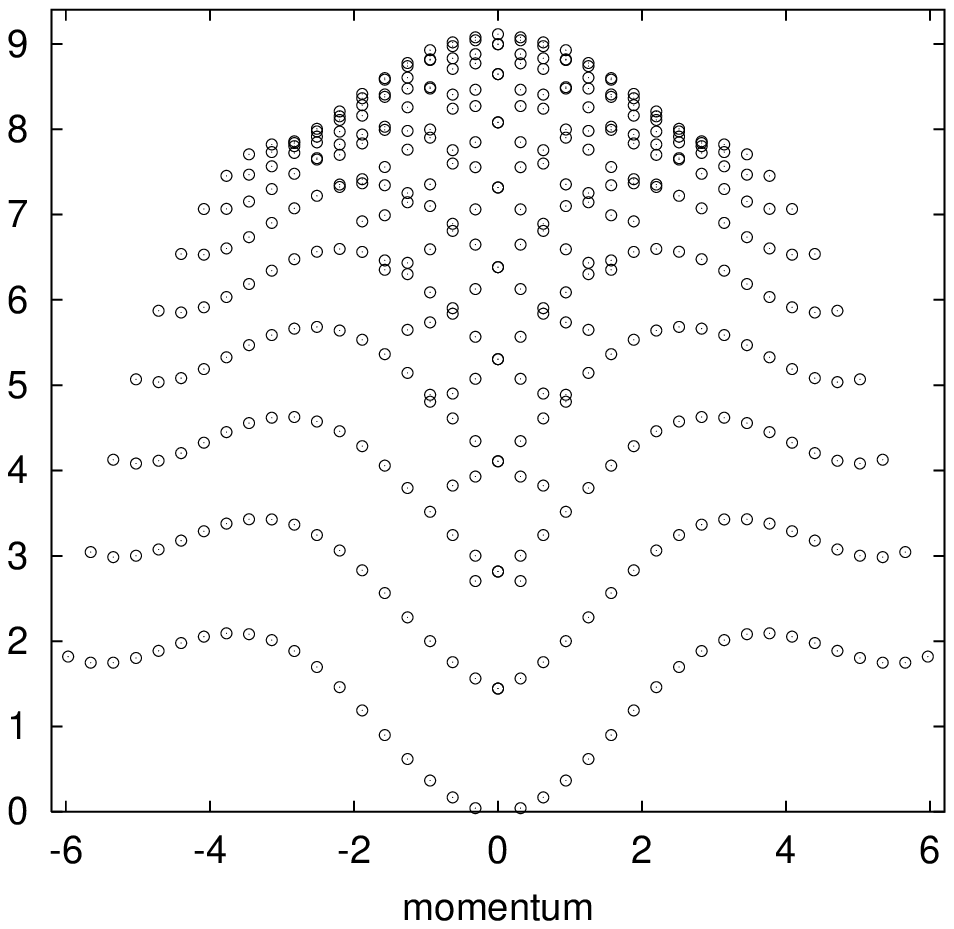}}}
\end{picture}
\vspace{0mm}
\caption{The holon-antiholon excitation spectrum (left)
and holon-isospinon excitation spectrum (right) calculated for L=20, N=20 
and $c=10$.}
\label{fig:spectra}
\end{figure}

For $N$ particles, the ground state is characterized by a one-row
$N$-column Young tableau $[ N ]$, of which the quantum numbers are
$\{ I_j\}:=\{-(N-1)/2,...,(N-1)/2\}$ and $\{J_\gamma\}= empty$. 
For this state eq. (\ref{eq:logBAE}) reduces to the case of \cite{LiebL}.
Here it is a $(N+1)$ fold multiplet with ${\cal I}^2=N(N+2)/4$.
The density of states per length of the ground state 
is plotted (fig.\ref{fig:density} left)
for various couplings. 
The ``particle''-hole (or called holon-antiholon) excitation
is defined by the quantum numbers:
$I_1 = -(N-1)/2+\delta_{1,j_1}$
(for $1\leq j_1\leq N$), 
$I_j = I_{j-1}+1+\delta_{j,j_1}$
(for $j=2,...,N-1$),
and $|I_N|\geq(N+1)/2$.
Fig. (\ref{fig:spectra}) the left is the corresponding excitation spectrum.
The isospinon-holon excitation is characterized by the Young tableau
$[N-1, 1]$, i.e., $M=1$.
Comparing to that of ground state,
the quantum number $\{I_j\}$ changes from
half-integer to integer or vice versa, accordingly, 
$I_1 = -N/2+\delta_{1,j_1}$ (for $1\leq j_1\leq N+1$),
$I_j = I_{j-1}+1+\delta_{j,j_1}$ (for $j=2,...,N$),
while $J_1=I_1+n$ so that $I_1<J_1<I_N$.
This is a $N-1$ fold multiplet with ${\cal I}^2=N(N-2)/4$. 
The excitation spectrum is plotted in Fig. (\ref{fig:spectra}). 
Two branches of the quasi-particle excitation was recently shown for
two-component condensate\cite{Goldstein} by means of the techniques
in nonlinear optics.
The density of state for $J_1=0$, $j_1=1$ is  plotted in Fig. 
(\ref{fig:density}). By comparison to the ground state where no
isospin rapidity exists, a rift emerged at the position
of the existing isospin rapidity for small $c$ that
is consistent to our previous analysis for weak coupling limit.            

In the thermodynamics limit, 
the Bethe-ansatz equations give rise to the following integral equations
for the density of roots and holes respectively in quasi-momentum 
and isospin rapidity spaces:
\begin{eqnarray}
\rho(k)+\rho_h(k) 
  &=& \frac{1}{2\pi}+\int_{Q}^{Q}dk'\rho(k')K_1(k-k')\nonumber\\
  \,&-&\int_{-B}^{B}d\lambda'\sigma(\lambda')K_{1/2}(k-\lambda'),
           \nonumber\\
\sigma(\lambda)+ \sigma_h(\lambda) 
  &=& \int_{-Q}^{Q}dk'\rho(k')K_{1/2}(\lambda-k')\nonumber\\
  \,&-&\int_{-B}^{B}d\lambda'\sigma(\lambda')K_1(\lambda-\lambda'),
\label{eq:thermodynamics}
\end{eqnarray}
where $K_\mu(x)=\pi^{-1}\mu c/(\mu^2 c^2+x^2)$. The $Q$ and $B$ are determined
consistently by 
$\int_{-Q}^{Q}\rho(k)dk=N/L$ and $\int_{-B}^{B}\sigma(\lambda)d\lambda=M/L$.
It is easy to check by Fourier transform that the state of
$B=\infty$ and $\sigma_h=0$ is an isospin singlet, but further
calculations show that it is not the ground state. The real ground 
state, however, corresponds to $\sigma=\rho_h=0$ in
eq. (\ref{eq:thermodynamics}),
which concludes that the ground state is an isospin ``ferromagnetic'' state
which agrees with the result of mean field theory\cite{Janson}.
Two-particle case is a pedagogical example to understand it.
For the two-body Schr\"odinger equation in the center-of-mass frame, the 
permutation of particle coordinates becomes the parity reflection of their
relative coordinate. The oscillation theorem in quantum mechanics tells that
the spatial wave function without zero nods 
which is also an even parity solution yields the lowest energy.
If the particles possess a SU(2) intrinsic degree of freedom, their intrinsic
wave function must be symmetric (anti-symmetric) to keep the total 
wave function
with the lowest energy being symmetric (ant-symmetric).
Then we may easily understand that the ground state of 
Bose system (Fermi system) is of ``ferromagnetic'' (anti-ferromagnetic). 

The highly degenerate ferromagnetic ground state which is merely true for the
vanished Zeemann term will suddenly slit up into Zeemann sublevels
once the external field is applied. 
The ground state whence becomes a polarized state 
when the Rabi field which breaks the SU(2) symmetry is turned on. 

In order to evaluate excitation energy we let
$\rho(k)=\rho_0(k)+\rho_1(k)/L$ ($\rho_0$ refers to ground state). 
In the presence of isospin degree of freedom, there will be holon-isospinon
excitation in addition to the holon-antiholon
excitation. The later is created by a hole inside
the quasi Fermi sea $\bar{k}\in[-k_F, k_F]$ and an additional 
$k_p$ outside it,
\[
\rho_1(k)+\delta(k-\bar{k})
  =\int_{-k_F}^{k_F}dk'\rho_1 K_1(k-k')+K_1(k-k_p).
\]
The excitation energy consists of two terms:
$\Delta E=\int k^2\rho dk +k_p^2=\varepsilon_h(\bar{k})+\varepsilon_a(k_p)$,
the holon energy $\varepsilon_h$ and antiholon energy 
$\varepsilon_a(k_p)=-\varepsilon_h(k_p)$ are given by,
\begin{eqnarray}
\varepsilon_h(y)&=&-y^2+\int_{-k_F}^{k_F}k^2\rho^h_1(k,y)dk,
   \nonumber\\
\rho^h_1(k,y)&+&K_1(k-y)=\int_{-k_F}^{k_F}dk'K_1(k-k')\rho^h_1(k',y).
\label{eq:holon}
\end{eqnarray}
Flipping one isospin corresponds to adding one isospin rapidity to the 
background of ``ferromagnetic'' ground state which brings about one hole 
in the $k$-sector inevitably.
The excitation energy $\Delta E =\int k^2\rho_1 dk$ is solved from
\[
\rho_1(k)+\delta(k-\bar{k})=\int_{-k_F}^{k_F}dk'K_1(k-k')\rho_1(k')
  -K_{1/2}(k-\lambda),
\]
consequently, $\Delta E=\varepsilon_h(\bar{k})+\varepsilon_{i}(\lambda)$.
The $\varepsilon_h$ is given by eq. (\ref{eq:holon}) and $\varepsilon_{i}$ by
$\varepsilon_{i}(\lambda)=\int k^2\rho^{i}_1(k,\lambda)dk$ with
\[
\rho^{i}_1(k,\lambda)+K_{1/2}(k-\lambda)
 =\int_{-k_F}^{k_F}dk'K_1(k-k')\rho^{i}_1(k',\lambda).
\]

\begin{figure}
\epsfclipoff
\fboxsep=0pt
\setlength{\unitlength}{0.8mm}
\begin{picture}(76,42)(0,0)
\linethickness{1pt}
\epsfysize=36mm
\put(-1,1){{\epsffile{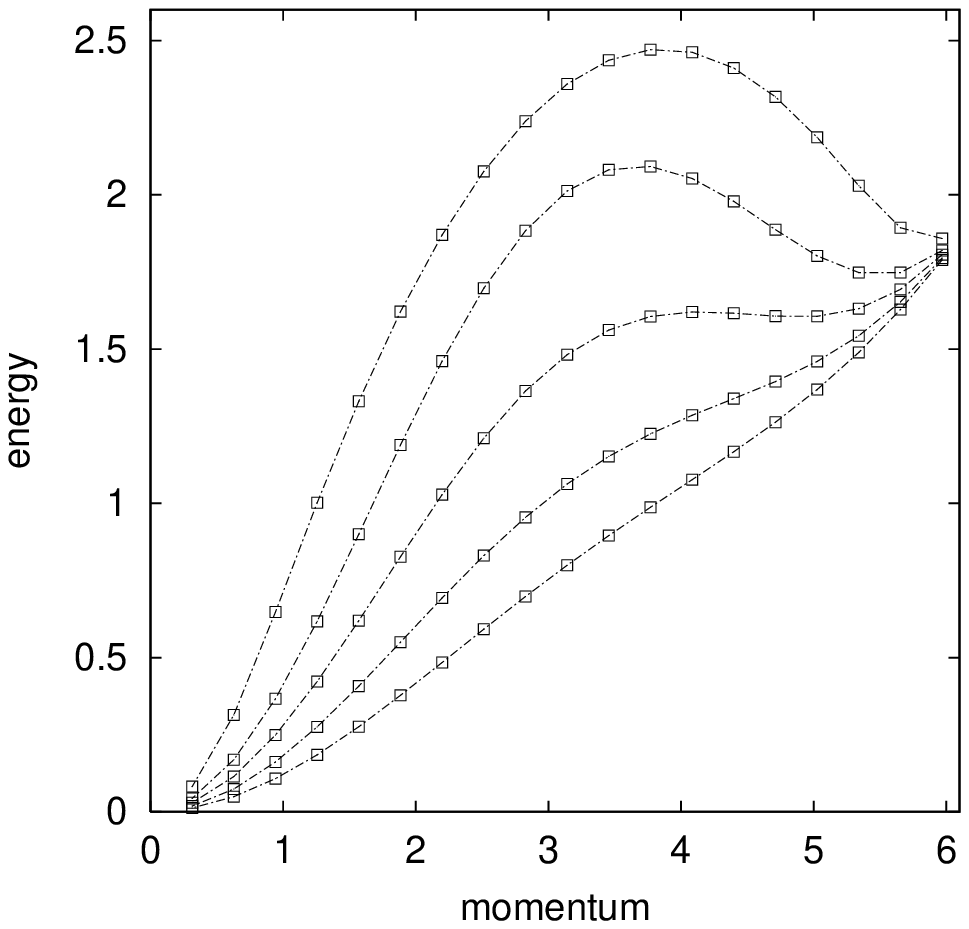}}}
\epsfysize=36mm
\put(50,1){{\epsffile{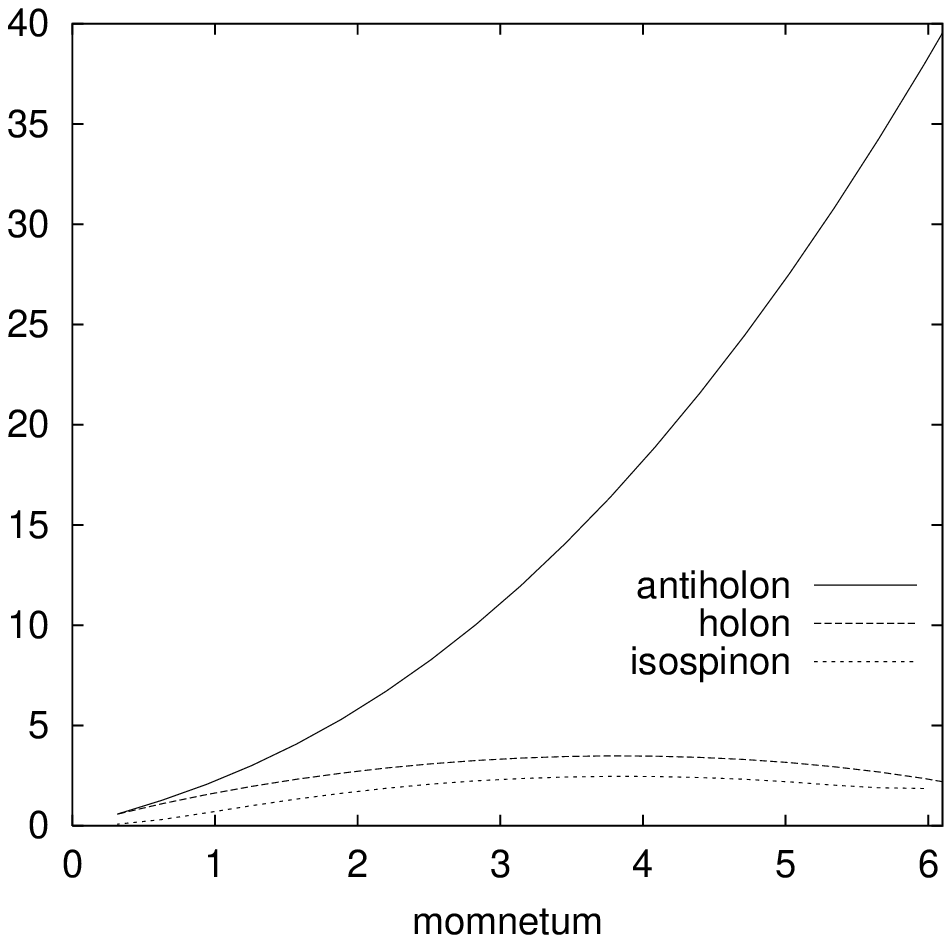}}}
\epsfysize=19mm
\put(54,20){{\epsffile{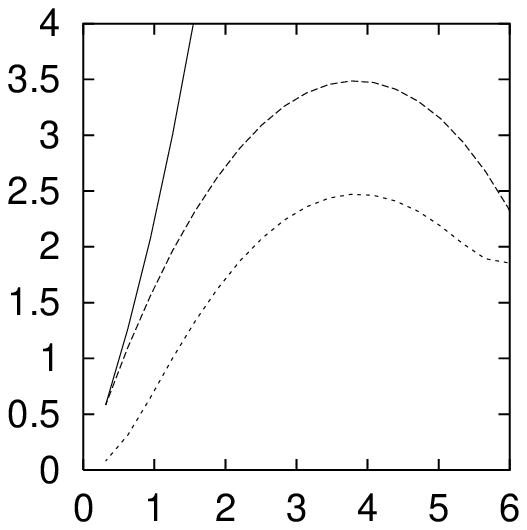}}}
\end{picture}
\vspace{0mm}
\caption{The left figure is the dispersion relations of isospinon
for different coupling constants where the curves from top to bottom
correspond to $c=1,10,20,40,80$ respectively.
The right figure is of antiholon, holon and isospinon for c=1.
They are calculated with $L=20$ and $N=20$.}
\label{fig:dispersins}
\end{figure}

In conclusion we found the existence of three elementary quasi-particles:
holon, antiholon and isospinon. From the asymptotic behaviors of those
basic modes as $\bar{k}$, $k_p$, and $\lambda$ tend to $k_F$, we find
that both the holon-antiholon and holon-isospinon excitations are gapless.
The related dispersions for
finite $N$ are plotted in Fig. (\ref{fig:dispersins}).
Differing from the spinons in a Fermi system,
the isospinon here is a triplet which
always couples to the ``ferromagnetic'' background antiparallelly.
Although it always accompanies charge excitation (holon) for single or 
odd number of isospinons,
the isospinons can be excited in pairs without exciting the U(1)
charge mode. 
Because of the coupling between
charge sector and isospin sector in eq.(\ref{eq:thermodynamics}), 
both cases brings about changes
in quasi-momentum distribution and hence turns out excitation energy.
The charge-isospin phase separation predicted by mean field 
theory \cite{phaseMFT} clearly occurs in present case due to the 
structure of eq.(\ref{eq:thermodynamics}). 
The holons and antiholons are quasi-particles
created in the momentum space while the isospinon behaves like 
dark soliton \cite{Carr} 
in the isospin sector that is tended to diminish an unit value from
the total isospin ${\cal I}$ eigenvalue.

Elaborating an experiment for two-component Bose gases whose 
transverse excitations are frozen out and
the dynamics becomes essentially one-dimensional\cite{Ketterle}
is expected to check the afore-mentioned
properties by the detection of the isospinon excitations
and the measurement of excitation spectra.

The work is supported by NSFC No.1-9975040, EYF and Trans-century projects of China
Ministry of Education.
YQL is also supported by AvH-Stiftung.
YQL thanks J.Cardy, J.Carmelo, H.Frahm, F.D.M. Haldane, 
T.C.Ho, M.Oshikawa and H.Saleur for interesting discussions;
also thanks F.Guinea for helpful remarks on the manuscript.

%

\end{document}